\documentclass[conference]{IEEEtran}
\IEEEoverridecommandlockouts

\usepackage{amsmath,amssymb,amsfonts,amsthm}
\usepackage{algorithmic}
\usepackage{graphicx}
\usepackage{textcomp}
\usepackage{xcolor}
\usepackage{orcidlink}
\usepackage{xcolor}
\usepackage{colortbl}
\usepackage{siunitx}
\usepackage[nolist]{acronym}
\usepackage{placeins}
\usepackage[caption=false,font=footnotesize]{subfig}
\usepackage{booktabs}
\usepackage{svg}
\usepackage[export]{adjustbox}
\usepackage{multirow}
\usepackage{float}
\usepackage{tabularx,booktabs, tabulary}
\usepackage{adjustbox}
\usepackage{graphicx}
\usepackage{svg}

\def\receiver{receiver}

\def\sender{sender}

\def\fieldRequirements{field-requirements}
\def\fieldRequirement{field-requirement}

\def\fieldRequirement{field-requirement}

\def\fieldRequirementProbabilities{field-requirement probabilities}

\def\FieldRequirementProbabilities{Field-requirement probabilities}

\def\unusedfields{unused fields}

\def\reduce{adapt}
\def\reduced{adapted}
\def\reduces{adapts}

\def\reducing{adapting}

\def\fieldmask{field mask}
\def\fieldmasks{field masks}

\def\messageTypes{message types}

\def\messageType{message type}

\usepackage[nolist]{acronym}
\usepackage{subfig}

\usepackage{listings}
\usepackage{svg}
\usepackage{siunitx}

\usepackage[normalem]{ulem}
\usepackage[style=ieee,dashed=false, maxcitenames=1,mincitenames=1,isbn=false,url=true,doi=false,date=year, hyperref=false]{biblatex}

\usepackage{amsmath,amssymb,amsfonts}
\usepackage{algorithmic}
\usepackage{graphicx}
\usepackage{textcomp}
\usepackage{xcolor}
\def\BibTeX{{\rm B\kern-.05em{\sc i\kern-.025em b}\kern-.08em
    T\kern-.1667em\lower.7ex\hbox{E}\kern-.125emX}}

\makeatletter
\def\ps@IEEEtitlepagestyle{%
  \def\@oddfoot{\mycopyrightnotice}%
  \def\@evenfoot{}%
}
\def\mycopyrightnotice{%
	\begin{minipage}{\textwidth}
		\centering \scriptsize
		© 2026 IEEE. Personal use of this material is permitted. Permission from IEEE must be obtained for all other uses, in any current or future media, including reprinting/republishing this material for advertising or promotional purposes, creating new collective works, for resale or redistribution to servers or lists, or reuse of any copyrighted component of this work in other works.\hfill
	\end{minipage}
	\gdef\mycopyrightnotice{}
}
\makeatother
    
\begin{document}

\title{Selective Field Transmission: Bandwidth Efficient Communication under Standardized Message Schemas\\
{}
}

\author{\IEEEauthorblockN{David Philipp Klüner}
\IEEEauthorblockA{\textit{Chair of Embedded Software} \\
\textit{RWTH Aachen University}\\
Aachen, Germany \\
kluener@embedded.rwth-aachen.de}
\and
\IEEEauthorblockN{David Murach}
\IEEEauthorblockA{\textit{Chair of Embedded Software} \\
\textit{RWTH Aachen University}\\
Aachen, Germany \\
david.murach@rwth-aachen.de}
\and
\IEEEauthorblockN{Stefan Kowalewski}
\IEEEauthorblockA{\textit{Chair of Embedded Software} \\
\textit{RWTH Aachen University}\\
Aachen, Germany \\
kowalewski@embedded.rwth-aachen.de}
\and
\IEEEauthorblockN{Alexandru Kampmann}
\IEEEauthorblockA{\textit{Chair of Embedded Software} \\
\textit{RWTH Aachen University}\\
Aachen, Germany \\
kampmann@embedded.rwth-aachen.de}
}

\maketitle

\begin{acronym}\itemsep0pt
    \acro{E/E}{Electrical/Electronic}
    \acro{IVN}{In-Vehicular Network}
    \acro{CAV}{Connected and Automated Vehicle}
    \acro{CAN}{Controller Area Network}
    \acro{AV}{Automated Vehicle}
    \acro{ARR}{Automatic Requirement Recording}
    \acro{AD}{Automated Driving}
    \acro{SDN}{Software-Defined Network}
    \acro{VM}{Virtual Machine}
    \acro{SOA}{Service-oriented Architecture}
    \acro{ROS 2}{Robot Operating System 2}
    \acro{CPS}{Cyber-Physical System}
    \acro{ECU}{Electronic Control Unit}
    \acro{AA}{Adaptive Application}
    \acro{IPC}{Inter-Process Communication}
    \acro{CM}{Communication Management}
    \acro{ROS}{Robot Operating System}
    \acro{RTPS}{Real-Time Publish-Subscribe Protocol}
    \acro{TLS}{Transport Security Layer}
    \acro{SM}{State Management}
    \acro{EM}{Execution Management}
    \acro{OEM}{Original Equipment Manufacturer}
    \acro{ARA}{AUTOSAR Runtime for Adaptive Applications}
    \acro{OS}{Operating System}
    \acro{FC}{Function Cluster}
    \acro{FPGA}{Field Programmable Gate Array}
    \acro{GPU}{Graphics Processing Unit}
    \acro{UCM}{Update and Configuration Management}
    \acro{DL}{Deep Learning}
    \acro{OROT}{OneRequirementOneTopic}
    \acro{OJT}{OneJointTopic}
    \acro{AP}{Adaptive Platform}
    \acro{IDS}{Intrusion Detection System}
    \acro{S2S}{Service-to-Signal}
    \acro{MQTT}{Message Queuing Telemetry Transport}
    \acro{OTA}{Over-The-Air}
    \acro{TSN}{Time-sensitive Networking}
    \acro{UDP}{User Datagram Protocol}
    \acro{NUC}{Intel Next-Unit-of-Computing}
    \acro{TCP}{Transmission Control Protocol}
    \acro{OMG}{Object Management Group}
    \acro{QoS}{Quality of Service}
    \acro{DDS}{Data Distribution Service}
    \acro{IP}{Internet Protocol}
    \acro{IDL}{Interface Definition Language}
    \acro{JSON}{JavaScript Object Notation}
    \acro{SDV}{Software-Defined Vehicle}
    \acro{SFT}{Selective Field Transmission}
    \acro{STR}{Selective Transmission Request}
    \acro{DMA}{Direct Memory Access}
    \acro{HMI}{Human Machine Interface}
    \acro{ML}{Machine Learning}
    \acro{SPDP}{Simple Participant Discovery Protocol}
    \acro{SEDP}{Simple Endpoint Discovery Protocol}
    \acro{API}{Application Programming Interface}
    \acro{IMU}{Inertial Measurement Unit}
    \acro{UCM}{Update and Configuration Management}
    \acro{HPC}{High-Performance Computer}
    \acro{Wi-Fi}{Wireless Fiber}
    \acro{Nav2}{Navigation 2}
    \acro{4G}{4G}
    \acro{HMI}{Human-Machine-Interface}
    \acro{NIC}{Network Interface Card}
\end{acronym}

\begin{abstract}
In this paper, we introduce and evaluate \acf{SFT}, a middleware mechanism that decouples transmission content from statically defined \messageTypes\ in publish-subscribe systems.
Industrial and robotics developers often face a dilemma: They can follow established best practices and use standard message types, such as in the \ac{ROS 2} and COVESA projects, to benefit from reusable and interoperable interfaces, or they can introduce proprietary, project-specific \messageTypes\ tailored to receiver requirements to reduce bandwidth.
\ac{SFT} resolves this trade-off by dynamically adapting the transmitted message components to each receiver's actual needs while preserving unmodified standard interfaces.
Receivers declare or automatically derive the required message components, which are communicated to the publisher. The publisher then serializes and transmits only the required component subset per receiver with minimal developer intervention.
Our evaluation shows that \ac{SFT} achieves significant bandwidth reductions without measurable per-message latency overhead, with savings proportional to the number and size of unused fields.

Implementation available at \url{https://github.com/embedded-software-laboratory/SelectiveFieldTransmission}.
\end{abstract}


\section{Introduction}
Developers of publish-subscribe systems face a recurring trade-off when defining message interfaces.
They can adopt standard message types, such as those provided by the \ac{ROS 2} or COVESA ecosystems, to benefit from reusable and interoperable interfaces across suppliers and components \cite{malavolta_how_2020}.
Alternatively, they can define project-specific \messageTypes\ tailored to each receiver's needs, reducing bandwidth at the cost of interoperability and maintainability.

This dilemma arises from a fundamental design property of current middleware: \messageTypes\ are defined in an \ac{IDL} and compiled into type-support code before deployment \cite{henle_architecture_2022, kluner_automotive_2025}.
\acp{IDL} guarantee type safety but fix the wire format, serialization logic, and subscription interface of every topic at the time of compilation \cite{macenski_robot_2022}.
Once a message type is chosen, every subscriber receives the complete message, regardless of how many fields it actually uses.

\begin{figure}[tbp]
    \centering
    \includegraphics[width=0.49\textwidth]{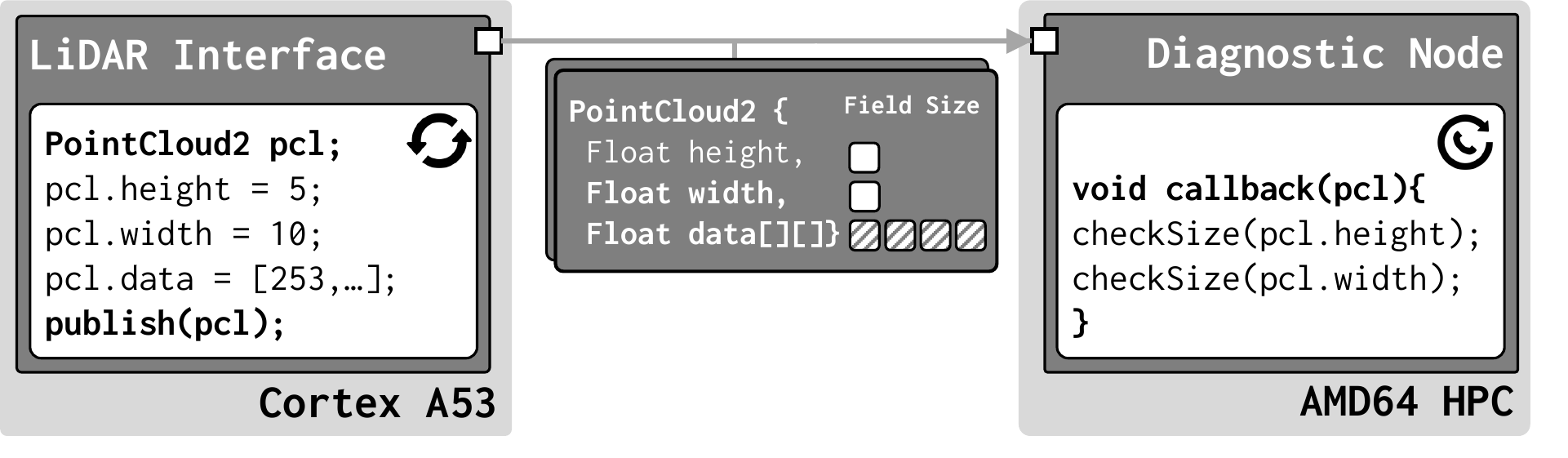}
    \caption{Motivating Example: Illustration of fields being unused and unnecessarily transmitted in a \ac{ROS 2} publish-subscribe system due to fixed message formats. Message derived from sensor\_msgs/msg/PointCloud2.}
    \label{fig:motivation}
\end{figure}

This rigidity is best illustrated by a motivating example, shown in Figure \ref{fig:motivation}.
Consider a software project in which a LiDAR supplier provides a perception node that publishes large \texttt{sensor\_msgs/PointCloud2} messages \cite{wu_oops_2021}.
During integration, a diagnostic node is added that monitors only the \texttt{height} and \texttt{width} fields of the published message to verify correct function.
Under current middleware design, this node must subscribe to the full message and deserialize the complete point cloud, including the raw sensor data, to access two scalar fields.
The alternative, optimizing this interface, requires asking the supplier to publish a second, narrower topic carrying only those fields.
This involves \ac{IDL} modification, code regeneration, node recompilation, and ongoing maintenance of a second topic, all to achieve narrower two-field access.
Should the diagnostic node later need an additional field, the entire cycle repeats.

These limitations are not confined to constructed scenarios, but recur in deployed \ac{ROS 2} software, such as the NAV2 navigation stack \cite{macenski_marathon_2020}.
The \texttt{nav2\_docking\_simple\_charging\_dock} node, discussed in detail in Section \ref{sec:eval:nav2}, requires a single field from the \texttt{BatteryState} message yet receives all 17.
These examples expose a core limitation: subscription granularity is limited to complete message types, forcing developers to choose between standard interfaces and efficient transmission.

This paper proposes \acf{SFT}, a middleware-level mechanism that resolves this trade-off by enabling field-level adaptation of transmitted messages without recompilation.
\ac{SFT} allows developers to retain existing \messageTypes\ while \reducing\ each transmitted message dynamically to only those fields actually requested by each receiver.
Subscribers declare their field requirements to publishers at run time, and publishers selectively serialize only the negotiated fields before transmission, eliminating unnecessary data at the source.
The mechanism is realized as an extension of the DDS middleware layer and integrated into the EmbeddedRTPS DDS stack.
The contributions of this paper are:
\begin{enumerate}
    \item The \acf{SFT} concept, comprising an explicit requirements API, an automatic recording mechanism for per-field requirements, a negotiation protocol, and a selective serialization path that encodes only the negotiated fields.
    \item An open-source implementation integrated into EmbeddedRTPS \footnote{https://github.com/embedded-software-laboratory/SelectiveFieldTransmission}.
    \item An evaluation of parametric and application-derived workloads quantifying bandwidth reduction and end-to-end latency impact.
\end{enumerate}

The remainder of this paper covers background (Section \ref{sec:background}), related work (Section \ref{sec:related}), the \ac{SFT} concept (Section \ref{sec:concept}), and evaluation (Section \ref{sec:eval}).

\section{Background}
\label{sec:background}

\paragraph{Communication Middleware}
\label{sec:background:middleware}
Communication middlewares are software frameworks between the \ac{OS} and user applications that enable inter-process and inter-device communication via IP-based protocols, exposing simplified \acp{API} for common patterns such as publish-subscribe \cite{wang_review_2024, kluner_automotive_2025, macenski_robot_2022}.
They typically incorporate runtime discovery, \acf{QoS} policies, and security features \cite{bode_systematic_2023}.
\ac{DDS}, with its \ac{RTPS} wire-protocol, is one such standard widely used in the industrial and robotics domains \cite{eprosima_14_2025}.

\begin{figure*}[tbp]
    \centering
    \includegraphics[width=0.9\textwidth]{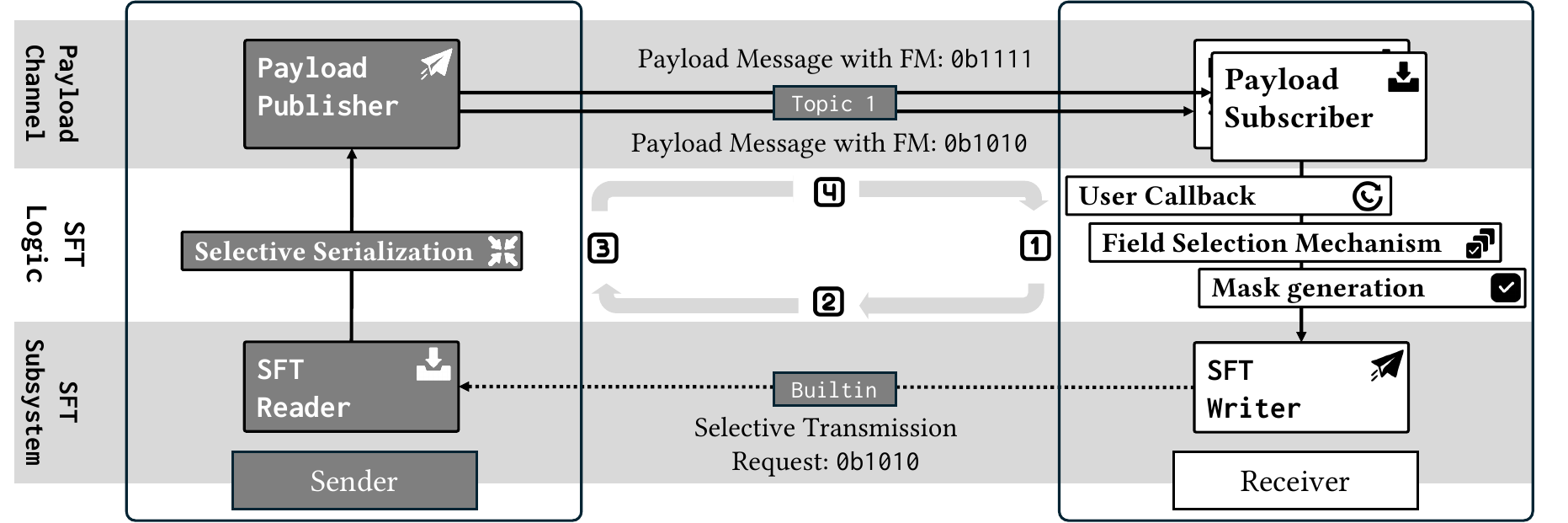}
    \caption{Simplified illustration of the components in our selective field transmission approach. Core components are enumerated in the middle of the figure and explained in order in Section \ref{sec:concept}.}
    \label{fig:aufbau}
\end{figure*}

\paragraph{Serialization}
For inter-device communication, middlewares transmit serialized byte data in a wire-format to ensure correct interpretation despite differences in endianness, alignment, or compiler behavior \cite{wolnikowski_zerializer_2021}.
\ac{CPS} middlewares typically rely on a pre-known structure rather than self-describing formats: code generated from \ac{IDL} files serializes and deserializes according to the message schema, using frameworks such as FlatBuffers or MicroCDR \cite{google_flatbuffers_2025}.

\section{Related Work}
\label{sec:related}
Existing work explores tailoring transmissions to receiver demands in several domains, although with mechanisms and assumptions that differ from our approach. 

\citeauthor{sperling_reducing_2024} \cite{sperling_reducing_2024} propose a request–response mechanism that transmits only regions-of-interest (ROIs) of an image rather than the full frame. 
Because ROIs typically cover a small fraction of the original image, the method substantially reduces bandwidth and end-to-end latency. 
In contrast, our method applies to arbitrary \messageTypes\ and does not depend on application-specific preprocessing.
\citeauthor{peeck_middleware_2021} \cite{peeck_middleware_2021} present a middleware protocol for efficient and reliable transmission of large multi-frame data samples over wireless links. The protocol maintains sample-level synchronization, uses frame-level retransmissions, and schedules transmissions according to timing constraints. 
This design is domain-specific and optimized for wireless environments with strict deadlines, whereas our approach targets general structured \messageTypes\ independent of the transport layer.
In the web domain, GraphQL, a query language and runtime for web APIs, provides clients with fine-grained control over which fields a server returns \cite{quina-mera_graphql_2023}. 
GraphQL is conceptually related in that clients declare required fields and servers return only those fields, paralleling \ac{SFT}'s receiver-driven approach.
However, GraphQL operates over HTTP in request-response mode, lacks real-time guarantees, and assumes schema introspection at query time, making direct adoption infeasible in \ac{CPS} publish-subscribe systems.
DDS-XTypes provide dynamic message construction at runtime through a type-builder API and automatic type discovery \cite{eprosima_14_2025}.
While XTypes' TypeObject negotiation can establish type compatibility automatically, reducing transmitted data still requires the developer to define, register, and match a narrower type to a dedicated topic for each distinct set of receiver requirements. 
Moreover, each XTypes-derived type requires its own topic and writer, incurring discovery and resource costs that scale with the number of distinct receiver requirements.
Consequently, for $n$ receivers with distinct \fieldRequirements, up to $n$ additional topics and writers must be maintained, each incurring discovery overhead and requiring recompilation whenever requirements change.
In contrast, \ac{SFT} serves all receivers from a single topic and automates the process from determining field requirements to constructing \reduced\ messages at serialization time, without requiring developer-managed type definitions.

\section{Selective Field Transmission}
\label{sec:concept}
\acl{SFT} introduces a feedback mechanism that enables publishers to \reduce\ outgoing messages to the actual needs of receivers at runtime. 
We implement \ac{SFT} as a DDS extension within the EmbeddedRTPS implementation \cite{kampmann_portable_2019}, but the concept is in principle applicable to any publish-subscribe system that provides access to the serialization layer and per-receiver dispatch at the transport level.
\acl{SFT} consists of four components, illustrated in Fig. \ref{fig:aufbau}, which together establish and use receiver-side field requirements to decouple transmitted data from \messageTypes:
\begin{enumerate}
    \item \textbf{Field Requirement Mechanism}: 
    \ac{SFT} first determines which fields of a message are actually used by each receiver. 
    These per-receiver \textit{\fieldRequirements} form the basis of the approach. 
    In Fig. \ref{fig:aufbau}, this step is shown on the receiver side after the user callback.
    We provide two mechanisms to obtain them, which are described in Section \ref{sec:concept:field-requirements}.
    \item \textbf{\ac{SFT} Channel}: 
    Once \fieldRequirements\ are known, they must be communicated to the \sender. 
    \ac{SFT} introduces new builtin readers and writers through which receivers send compact requests to the \sender. 
    In Fig. \ref{fig:aufbau}, this channel spans the illustration from the \ac{SFT} publisher to the \ac{SFT} subscriber and is detailed in Section \ref{sec:concept:request-channel}.
    \item \textbf{Selective Serialization}:
    Given the incoming \fieldRequirements\, the publisher determines how communication should be adapted. 
    \ac{SFT} \reduces\ the messages by omitting \unusedfields\ during serialization for each receiver. 
    This is illustrated in Fig. \ref{fig:aufbau} on the sender side within the SFT logic.
    For this selective serialization, we extend the underlying serialization toolchain, which is described in Section \ref{sec:concept:msg-reduction}.
   \item \textbf{Targeted Transmission}:
    The per-receiver \reduced\ messages must then be distributed to their respective receivers. 
    To accomplish this efficiently, we use a single topic with per-receiver message variants, discussed in Section \ref{sec:concept:targeted-transmission}.
\end{enumerate}

In the following sections, we discuss how we realize these components in order.

\subsection{Field Requirement Mechanism}
\label{sec:concept:field-requirements}
\ac{SFT} requires knowledge of which fields in a message are actually used by each receiver.  
For message type $m \in M$, we denote the set of fields that are accessed by receiver $r \in R$ as $A_{r,m} \subseteq F_m$, where $F_m = \{1,\dots,K_m\}$ is the index set of all $K_m$ fields in message type $m$. 
\ac{SFT} provides two complementary mechanisms to establish these receiver-side \fieldRequirements.

\paragraph{Requirements API}
The explicit requirements API allows user applications to define the \fieldRequirements\ $A_{r,m}$ for a message $m$ and receiver $r$ directly.
The developer specifies the \fieldmask\ representation of requirements needed by the receiver. \ac{SFT} encodes this set into a \ac{STR} without relying on runtime observation, thereby guaranteeing correctness, repeatability, and independence from runtime variability.
We represent \fieldRequirements\ as a bit vector, referred to as the \textit{\fieldmask}, in which each bit corresponds to a field and indicates whether it is required.
We refer to \fieldmasks\ where every field is required as \textit{full-mask} requirements.

\begin{figure}[tbp]
    \centering
    \includegraphics[width=0.49\textwidth]{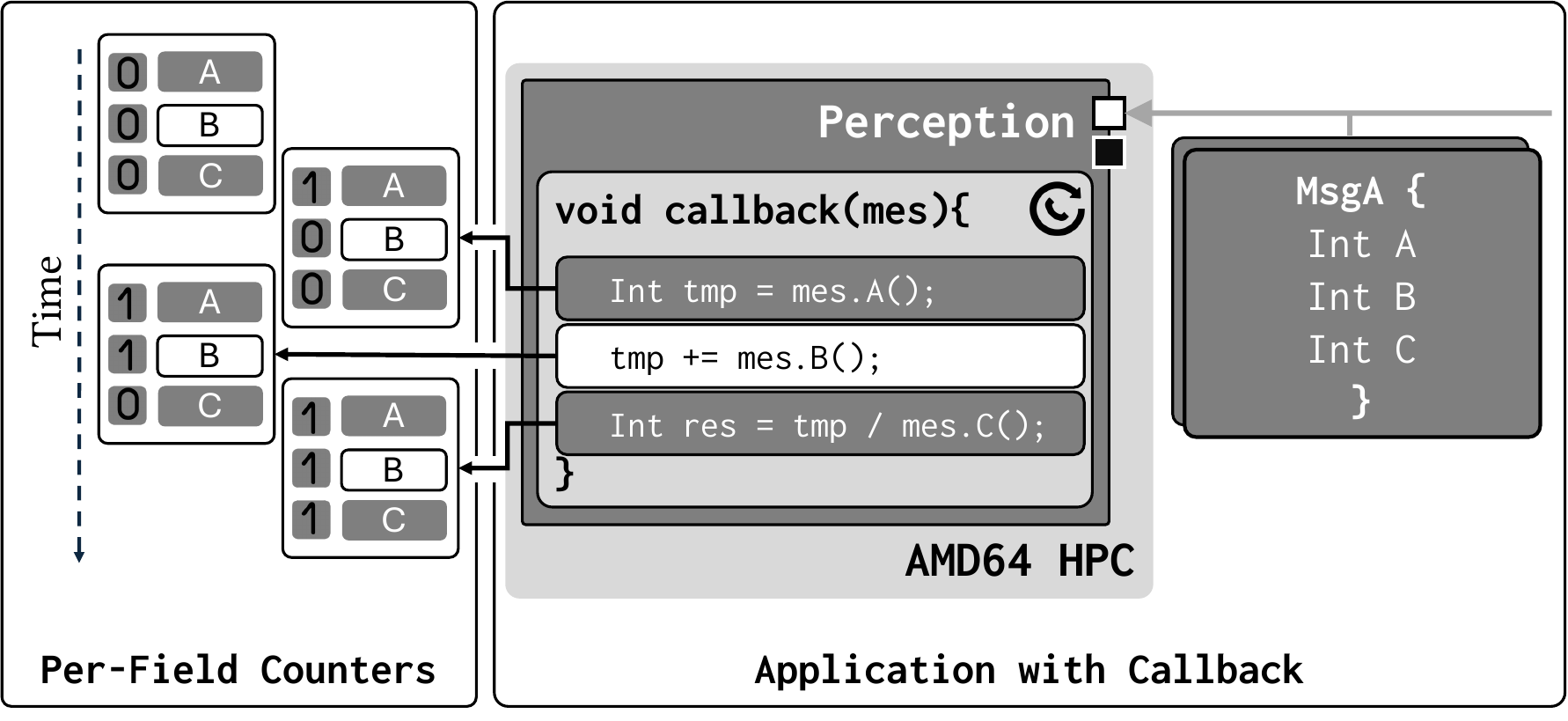}
    \caption{Illustration of Automatic Requirement Recording. Per-field counters are incremented at runtime when a \texttt{function} is called in the callback to the user-application.}
    \label{fig:ARR}
\end{figure}

\paragraph{Automatic Requirement Recording (ARR)}
As a user-friendly alternative to explicit specification, ARR determines \fieldRequirements\ $A_{r,m}$ automatically at runtime by recording which fields an application accesses while processing messages. Fig. \ref{fig:ARR} illustrates the method using an example message.

When a message is received for the first time, the \ac{SFT} layer first receives the full message from DDS and forwards it to the user application’s callback function.
Field accesses within the callback are performed via generated \texttt{getter} functions.
Our modified serialization code-generation tool-chain instruments these functions to record access counts for each field.
The generated structures contain counters for each field, which are not part of the wire-representation and incremented by each call to a corresponding \texttt{getter} function.
After each callback, \ac{SFT} extracts the set of accessed fields as a \fieldmask.
In our implementation, readers track per-field access counters across callback invocations and automatically re-announce their \fieldRequirements\ via the \ac{SFT} channel whenever the derived \fieldmask\ changes.

As ARR reflects only observed behavior, it may underestimate requirements when conditional branches access fields not yet triggered at runtime.
In such cases, the receiver may currently obtain default-initialized values (zero for numeric types, empty for sequences) for unrequested fields.
This behavior places correctness responsibility on the requirement mechanism.
The explicit requirements API guarantees complete \fieldRequirements\ independent of runtime observations and is therefore recommended for most deployments.
ARR is best suited for scenarios where field access patterns are stable and where ease-of-use is important.
We plan to add a retransmission mechanism to ARR in future work that allows receivers to detect and recover missing fields.

\subsection{SFT Channel}
\label{sec:concept:request-channel}
To communicate the \fieldRequirements\ $A_{r,m}$ from receivers $r \in R$ to the sender, \ac{SFT} introduces a secondary feedback channel using new builtin reader and writers, shown in the lower third of Fig. \ref{fig:aufbau}. 

\begin{figure*}[tbp]
    \centering
    \includegraphics[width=0.85\textwidth]{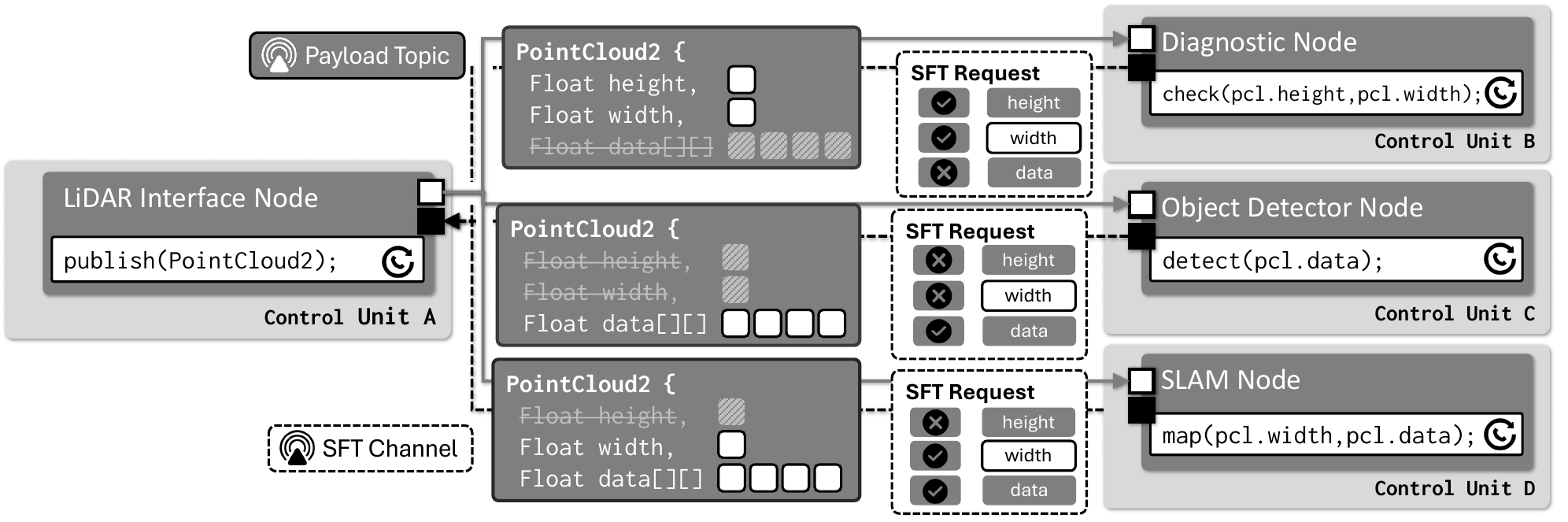}
    \caption{Example system derived from the motivating example. A LiDAR publisher sends \texttt{PointCloud2} messages to a diagnostic node (requiring \texttt{height}, \texttt{width}), an object detector (requiring \texttt{data}), and a SLAM node (requiring all fields except \texttt{height}). Each receiver emits an \ac{STR} (dashed boxes, center right). The sender \reduces\ each transmission accordingly. Grey fields are omitted, white fields are transmitted.}
    \label{fig:example}
\end{figure*}

On the receiver side, an \ac{SFT} writer emits \acp{STR} encoding the current \fieldRequirements\ as a \fieldmask.
On the sender side, an \ac{SFT} reader collects these requests and maintains a per-reader \fieldmask, which determines how outgoing messages are serialized for each receiver.
Figure \ref{fig:example} illustrates the \acp{STR} received from each receiver within a larger example system.
Whenever a \receiver\ detects that its \fieldRequirements\ have changed, i.e., the updated set of accessed fields $A_{r,m}' \neq A_{r,m}$, or when a new \receiver\ joins, it sends a new \ac{STR}.
Until a receiver's requirements are known, the sender transmits full messages containing all fields \(F_m\) to that receiver.

In our implementation, the \ac{SFT} channel is realized through dedicated builtin readers and writers that carry a message containing the \fieldmask\ and a monotonic sequence number.
We announce \ac{SFT} endpoint presence using a vendor-specific flag in the \ac{SPDP} message, so that the additional endpoints incur no extra discovery overhead.

\subsection{Selective Serialization}
\label{sec:concept:msg-reduction}
Once receiver \fieldRequirements\ have been exchanged, messages can be \reduced\ according to the per-reader \fieldRequirements\ $A_{r,m}$ for each receiver $r \in R$ and \messageType\ \(m \in M\).
We implement this reduction through a modified serialization toolchain that introduces serialization functions accepting a \fieldmask\ provided by \ac{SFT}.
The toolchain is based on MicroCDR XRCE Gen and MicroCDR.
Using these functions, the \ac{SFT} \sender\ omits all fields \(j \in F_m \setminus A_{r,m}\) by skipping their serialization primitives, and serializes only the required subset.
The resulting wire-format message therefore contains exactly the fields permitted by the mask.
For a receiver $r$ and \messageType\ $m$ with field sizes $s_{m,j}$, the baseline transmits $B_m = \sum_{j \in F_m} s_{m,j}$ bytes per message, while \ac{SFT} transmits only
$B_{r,m}^{\text{SFT}} = \sum_{j \in A_{r,m}} s_{m,j},$
yielding per-receiver savings of $\Delta_{r,m} = \sum_{j \in F_m \setminus A_{r,m}} s_{m,j}$.
For variable-length fields such as arrays, $s_{m,j}$ denotes the actual serialized size of field $j$ in a given message instance. 
Savings from omitting such fields therefore vary per message.
When multiple receivers share the same \fieldmask, the sender reuses the same serialized variant. Otherwise, it produces distinct variants for each unique mask.
In our implementation, each writer maintains per-reader \fieldmasks\ updated via the \ac{SFT} channel and serializes outgoing messages individually per reader proxy, defaulting to all fields until requirements are received.
A dedicated history cache creates and caches per-mask serialized message variants.

\subsection{Targeted Transmission}
\label{sec:concept:targeted-transmission}
Once serialized variants have been produced, the \reduced\ messages are transmitted on a single topic using unicast transmissions.
The \ac{SFT} sender stores the current \fieldmask\ $A_{r,m}$ for each reader $r$ and, upon transmission, selects the corresponding serialized message variant for each receiver.
To transmit the active \fieldmask\ to the receiving side without additional communication overhead, \ac{SFT} encodes it into the often-unused RTPS header fields for \texttt{inlineQoS} and \texttt{extraFlags}.
Figure \ref{fig:example} illustrates the separate message versions created for each receiver within a larger example system.
On reception, the reader extracts this bitmask and performs partial deserialization, reconstructing only the fields in $A_{r,m}$.
Fields not included in $A_{r,m}$ are default-initialized on the receiver side. Correctness guarantees depend on the chosen requirement mechanism, as discussed in Section \ref{sec:concept:field-requirements}.
The current implementation uses a \fieldmask\ representation of 32 bits, supporting up to 32 top-level fields. This limit can be extended as needed. \ac{SFT} currently operates at the granularity of top-level fields only.

\subsection{Example}
\label{sec:concept:example}
Figure \ref{fig:example} illustrates \ac{SFT} applied to an expanded version of the motivating example with three receivers of varying \fieldRequirements.
The \texttt{PointCloud2} message type has $K_m = 3$ top-level fields, with the \texttt{data} array dominating message size at over \SI{16}{\kilo\byte}.
The diagnostic node requires only two scalar fields, $A_{\text{diag},m} = \{\texttt{height}, \texttt{width}\}$, the object detector requires only the point cloud, $A_{\text{det},m} = \{\texttt{data}\}$, and the SLAM node requires most fields except \texttt{height}, $A_{\text{slam},m} = F_m \setminus \{\texttt{height}\}$.
Since all three \fieldmasks\ are distinct, the sender produces three serialized variants.
The diagnostic node benefits most: omitting the large \texttt{data} field reduces $B_{\text{diag},m}^{\text{SFT}}$ from more than \SI{16}{\kilo\byte} to a few scalar fields, saving significant resources and potentially enabling resource-constrained devices to participate in communication that would otherwise exceed their capacity.
Savings for the object detector and SLAM node are comparably small, as the omitted fields contribute little to the overall message size relative to the \texttt{data} array.
In the baseline, all three receivers would each receive the full $B_m$ per message, while \ac{SFT} transmits only the required subset per receiver, reducing aggregate bandwidth roughly by one third. 
The evaluation in Section \ref{sec:eval:pc2} quantifies these reductions with measured data.

\section{Evaluation}
\label{sec:eval}
We evaluate our \ac{SFT} approach using three complementary methods.
Our core goal is to validate the \ac{SFT} concept and quantify its impact on bandwidth, latency, and resource usage:
\begin{enumerate}
\item \textbf{Parametric Study}: First, we evaluate system behavior using our implementation.
Systematic configurations allow structured measurements of latency, bandwidth, and resource usage.

\item \textbf{Nav2 Experiments}: Next, we assess \ac{SFT} in robotics scenarios using workloads derived from the \ac{ROS 2} \ac{Nav2} stack. Because synthetic \messageTypes\ and \fieldRequirements\ may not fully capture real-world scenarios, we apply our method to three real message types within \ac{Nav2}-derived scenarios.

\item \textbf{PointCloud2 Case Study}: Finally, we evaluate \ac{SFT} against our motivating example (Fig.~\ref{fig:motivation} and \ref{fig:example}), demonstrating \ac{SFT}'s per-receiver adaptation with heterogeneous field sizes.
\end{enumerate}

All evaluation methods rely on a shared set of metrics, as well as parameters summarized in Section \ref{sec:eval:parameters}.
\begin{figure*}[t]
    \centering
    Parametric Study Bandwidth and Latency Measurements\par\medskip
    \includegraphics[width=0.95\textwidth]{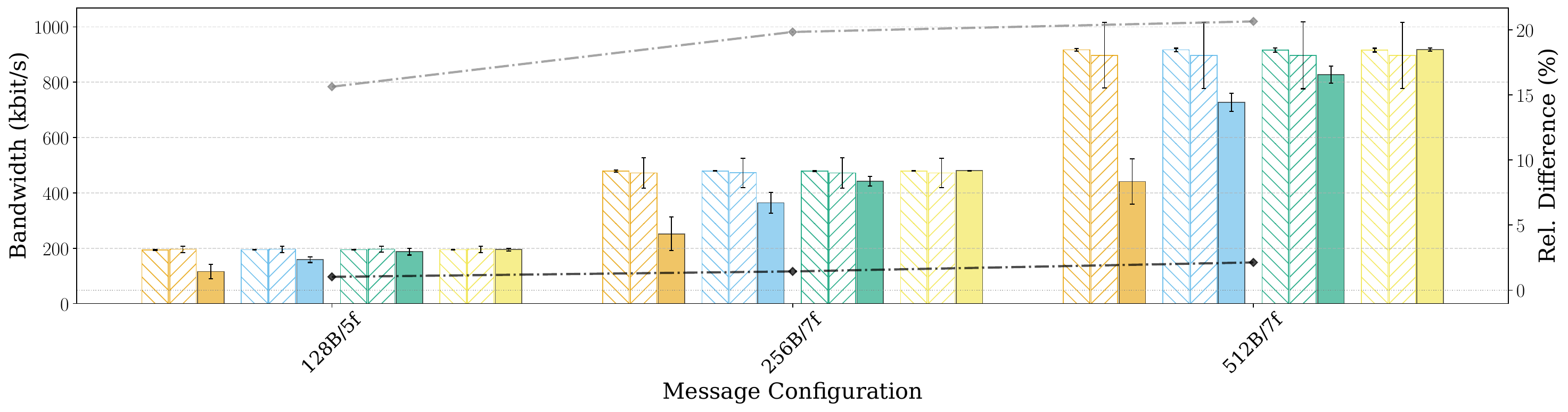}
    \includegraphics[width=0.95\textwidth]{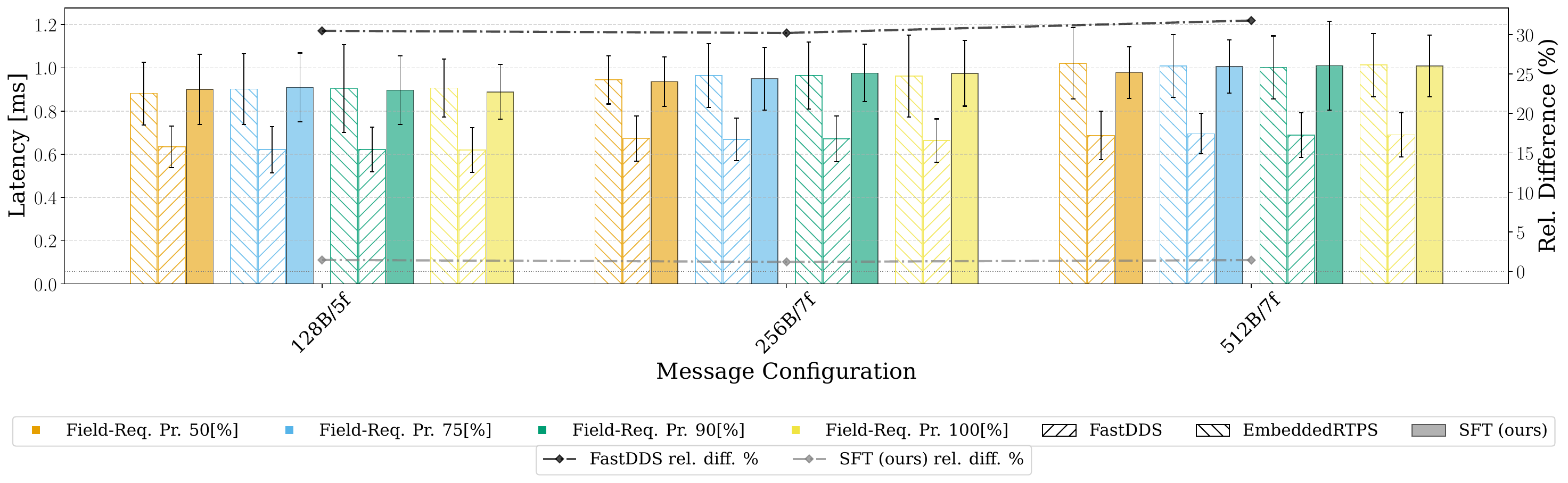}
    \caption{Overview of mean aggregate network bandwidth usage and mean communication latency for our parametric study. Results shown were measured in a topology with three receivers and one sender, with three different message types and four requirement probabilities following the notation presented in Section \ref{sec:eval:parameters}. Bars show means, error bars indicate one standard deviation.}
    \label{fig:results:param_study}
\end{figure*}
We measure transmission latency ($L_{T}$), defined as the time between \acl{SFT}'s \texttt{publish()} call and the receiver's user-application callback, network bandwidth ($B$) between communicating peers.

\subsection{Parameters}
\label{sec:eval:parameters}
The parameters reflect the communication characteristics reported in robotics and industrial systems \cite{wu_oops_2021, kluner_automotive_2025}:
\begin{itemize}
    \item \textbf{Field Count} ($|F_m| \in \{5,7\}$): Number of fields per message, reflecting common robotics message structures \cite{wu_oops_2021}.
    \item \textbf{Field Size} ($s_{m,j} \in \{$\SIlist{128;256;512}{\byte}$\}$): Per-field size, uniform within each configuration.
    \item \textbf{Receiver Count} ($|R| \in \{1,3\}$): Topologies are denoted as \texttt{machines:senders\,-\,machines:receivers}, e.g., 1:1-3:3 is one sender transmitting to three receivers on separate machines.
    \item \textbf{\FieldRequirementProbabilities} ($FRP := P(j \in A_{r,m}) \in [0.5, 1.0]\,$): Per-field independent and identically distributed probability of being required. Each receiver independently draws its own mask from this distribution, so receivers within the same experiment typically operate under distinct \fieldRequirements\ served from a single topic.
\end{itemize}

\subsection{Experimental Setup}
We conducted our experiments in a testbed consisting of four Lenovo Thinkcenters M900, each with an Intel Core i5-6500T processor, 16 GB of DDR4 RAM, and an M.2 SSD. 
All machines were interconnected through a central switch using 1 Gbit full-duplex Ethernet. 
We fix the sending frequency to \SI{10}{\hertz} in all experiments to keep the parameter space tractable.
We use EmbeddedRTPS \cite{kampmann_portable_2019} and FastDDS 3.4 as the baseline DDS implementation and implement our proposed SFT extensions in an EmbeddedRTPS fork. 
We used Ubuntu 24.04.3 LTS with a (GNU/Linux 6.17.0-14-generic x86\_64) kernel on our hardware. We did not implement additional real-time configuration. Our experiments used TShark for bandwidth recordings, LTTng for tracing, and PTP to ensure time synchronization.

\begin{table}[tbp]
    \centering
    \caption{Parametric Study Comparison}
    \label{tab:eval:results_comparison}

    \begin{tabular}{lccc}
    \toprule
    \textbf{Metric} & \textbf{FastDDS} & \textbf{EmbeddedRTPS} & \textbf{SFT (ours)} \\
    \midrule
    \multicolumn{4}{l}{\textit{Mean Bandwidth $B$ [\SI{}{\kilo\bit\per\second}]}} \\
    \hspace{1mm} All Exp.              & 521.8 (\SI{-1.5}{\percent})  & 530.0  & 425.7 (\textbf{\SI{-19.7}{\percent}}) \\
    \hspace{1mm} FRP \SI{50}{\percent} & 521.9 (\SI{-1.5}{\percent})  & 529.9  & 269.7 (\textbf{\SI{-49.1}{\percent}}) \\
    \hspace{1mm} FRP \SI{75}{\percent} & 521.9 (\SI{-1.6}{\percent})  & 530.2  & 416.7 (\textbf{\SI{-21.4}{\percent}}) \\
    \hspace{1mm} FRP \SI{90}{\percent} & 521.8 (\SI{-1.6}{\percent})  & 530.0  & 485.4 (\textbf{\SI{-8.4}{\percent}}) \\
    \hspace{1mm} FRP \SI{100}{\percent}& 521.8 (\SI{-1.5}{\percent})  & 529.9  & 531.0 (\textbf{\SI{+0.2}{\percent}}) \\
    \midrule
    \multicolumn{4}{l}{\textit{Mean Transmission Latency $L_T$ [\SI{}{\milli\second}]}} \\
    \hspace{1mm} All Exp.              & 0.66 (\SI{-30.9}{\percent})  & 0.96  & 0.95 (\textbf{\SI{-0.3}{\percent}}) \\
    \hspace{1mm} FRP \SI{50}{\percent} & 0.66 (\SI{-30.2}{\percent})  & 0.95  & 0.94 (\textbf{\SI{-1.5}{\percent}}) \\
    \hspace{1mm} FRP \SI{75}{\percent} & 0.66 (\SI{-30.9}{\percent})  & 0.96  & 0.95 (\textbf{\SI{-0.3}{\percent}}) \\
    \hspace{1mm} FRP \SI{90}{\percent} & 0.66 (\SI{-30.9}{\percent})  & 0.96  & 0.96 (\textbf{\SI{+0.4}{\percent}}) \\
    \hspace{1mm} FRP \SI{100}{\percent}& 0.66 (\SI{-31.4}{\percent})  & 0.96  & 0.96 (\textbf{\SI{-0.0}{\percent}}) \\
    \bottomrule
\end{tabular}
\end{table}

\subsection{Parametric Study}
\label{sec:eval:parametric}
We evaluated our implementation in 96 experiments in five repeat runs with all configurations derived from parameters outlined in Section \ref{sec:eval:parameters}. 
In these experiments, we used regular transmissions of the full payload using EmbeddedRTPS as the direct baseline, since \ac{SFT} extends it. FastDDS is included as an independent reference implementation. All experiments used \ac{ARR} to derive \fieldRequirements.
EmbeddedRTPS exhibits slightly higher latency than FastDDS, likely due to its internal thread-pool introducing coordination overhead on both send and receive paths. 
Experiments derived their field requirements i.i.d based on the given \fieldRequirementProbabilities.

\subsubsection{Results}
Our empirical results show that \ac{SFT} reduces bandwidth in nearly all configurations.
Fig. \ref{fig:results:param_study} and Table \ref{tab:eval:results_comparison} summarize the observed transmission latencies and bandwidths and report differences compared to baseline.
On average, \ac{SFT} reduced traffic across all experiments by \SI{19.7}{\percent}, with mean reductions reaching \SI{49.1}{\percent} at $FRP = \SI{50}{\percent}$.
Due to our middleware integration, \ac{SFT} reduced bandwidth whenever at least one field can be omitted, and in full-mask configurations the bandwidth overhead remained at most \SI{0.2}{\percent} compared to EmbeddedRTPS. 
The magnitude of the bandwidth reduction depends on field sizes and the specific \fieldRequirements\ of each receiver.
The \ac{Nav2} case study addresses this by using real message structures with heterogeneous field sizes.

\begin{figure}[tbp]
    \centering
    \centering
    Response Time Example\par\medskip
    \includegraphics[width=0.45\textwidth]{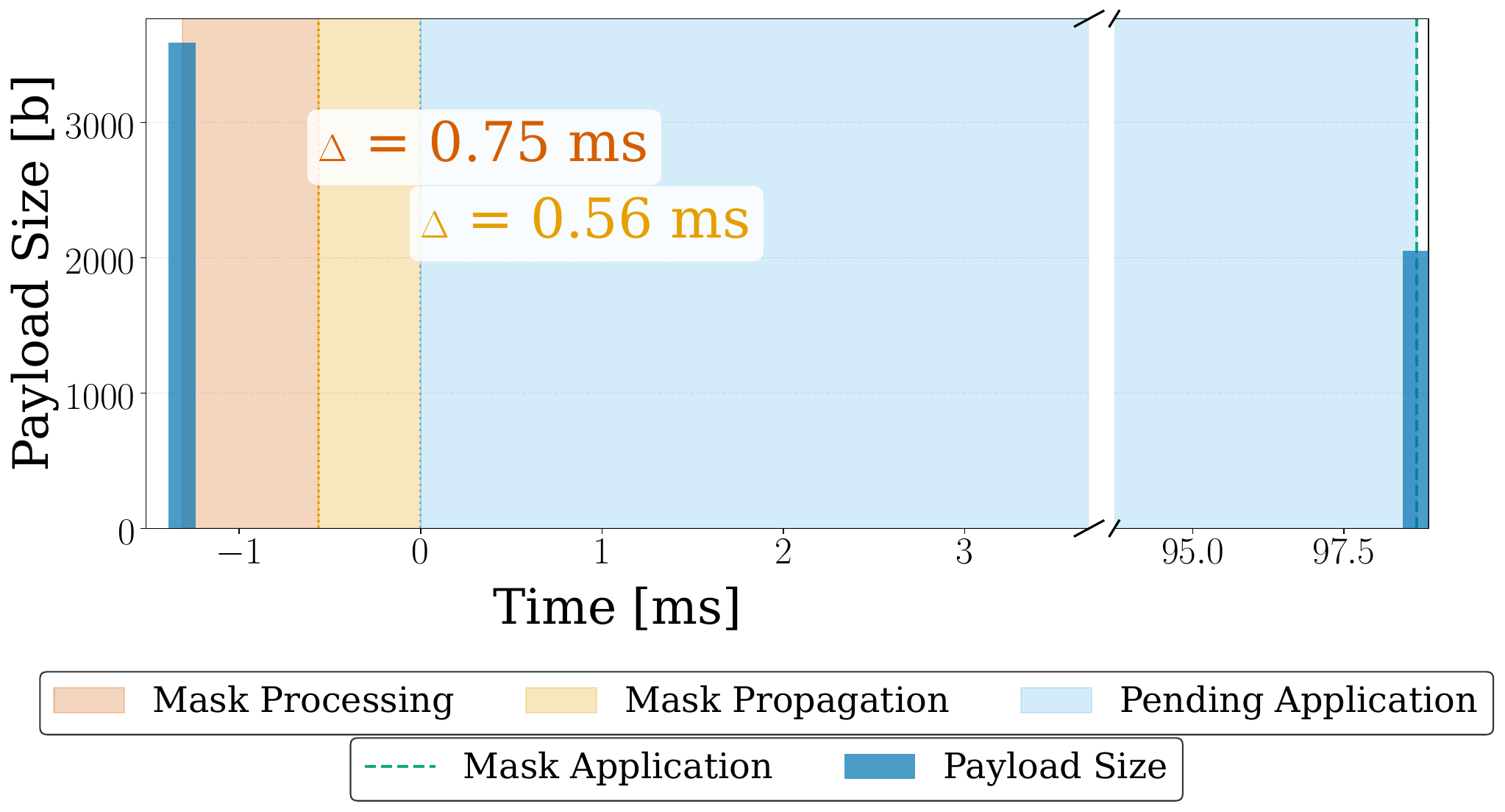}
    \caption{Representative example showing generation, transmission and application of a new field mask using ARR within an 512 byte 7 field and \SI{50}{\percent} FRP experiments. The next send payload message at 10Hz is directly reduced in size, demonstrating response times of \SI{1.3}{\milli\second} in this case.}
    \label{fig:results:response}
\end{figure}

In all experiments, \ac{SFT} introduced no measurable latency overhead compared to EmbeddedRTPS. 
Due to the middleware-layer implementation, only additional serialization overhead is required to create the \reduced\ messages for each receiver.
Regarding the method's response time, in our experiments a newly derived \fieldRequirement\ was emitted and received by the sender within \SI{1.3}{\milli\second} on average, taking effect on the next transmitted message.
Figure \ref{fig:results:response} illustrates one such response time (\SI{1.3}{\milli\second}) for an ARR-induced field requirement change.
CPU overhead increased by less than one percent in the mean compared to the EmbeddedRTPS baseline.
Memory usage increased on average by \SI{386.7}{\kilo\byte} over EmbeddedRTPS's \SI{39.4}{\mega\byte} baseline, a \SI{1.0}{\percent} increase.

\subsection{Nav2 Case Studies}
\label{sec:eval:nav2}
To complement the parametric study, we derived three case studies from real message types and field-usage patterns in the \ac{Nav2} stack \cite{macenski_marathon_2020}:

\begin{itemize}
    \item The \texttt{nav2\_amcl} \footnote{https://github.com/ros-planning/navigation2/blob/main/nav2\_amcl/src/\\amcl\_node.cpp} component implements an Adaptive Monte-Carlo Localizer in \ac{Nav2}. 
    It subscribes to the \texttt{LaserScan} message but does not use the large \texttt{intensities} array in the message which accounts for a large part of the message, all other fields are used. 

    \item The \texttt{simple\_charging\_dock} \footnote{https://github.com/ros-navigation/navigation2/blob/main/nav2\_docking/\\opennav\_docking/src/simple\_charging\_dock.cpp} component implements an example for \ac{Nav2}'s framework to auto-dock robots and exclusively evaluates the \texttt{current} field out of 17 fields in the \texttt{BatteryState} message.  
   
    \item \ac{Nav2} components, such as the \texttt{map\_saver} \footnote{https://github.com/ros-navigation/navigation2/blob/main/nav2\_map\_server\\/src/map\_saver/map\_saver.cpp}, implementing map saving features, rely on all fields of the \texttt{OccupancyGrid} message, including metadata and the entire occupancy data array. 
\end{itemize}

\subsubsection{Results}
\begin{figure}[tbp]
    \centering
    NAV2 Case Studies Bandwidth Measurement\par\medskip
    \includegraphics[width=0.45\textwidth]{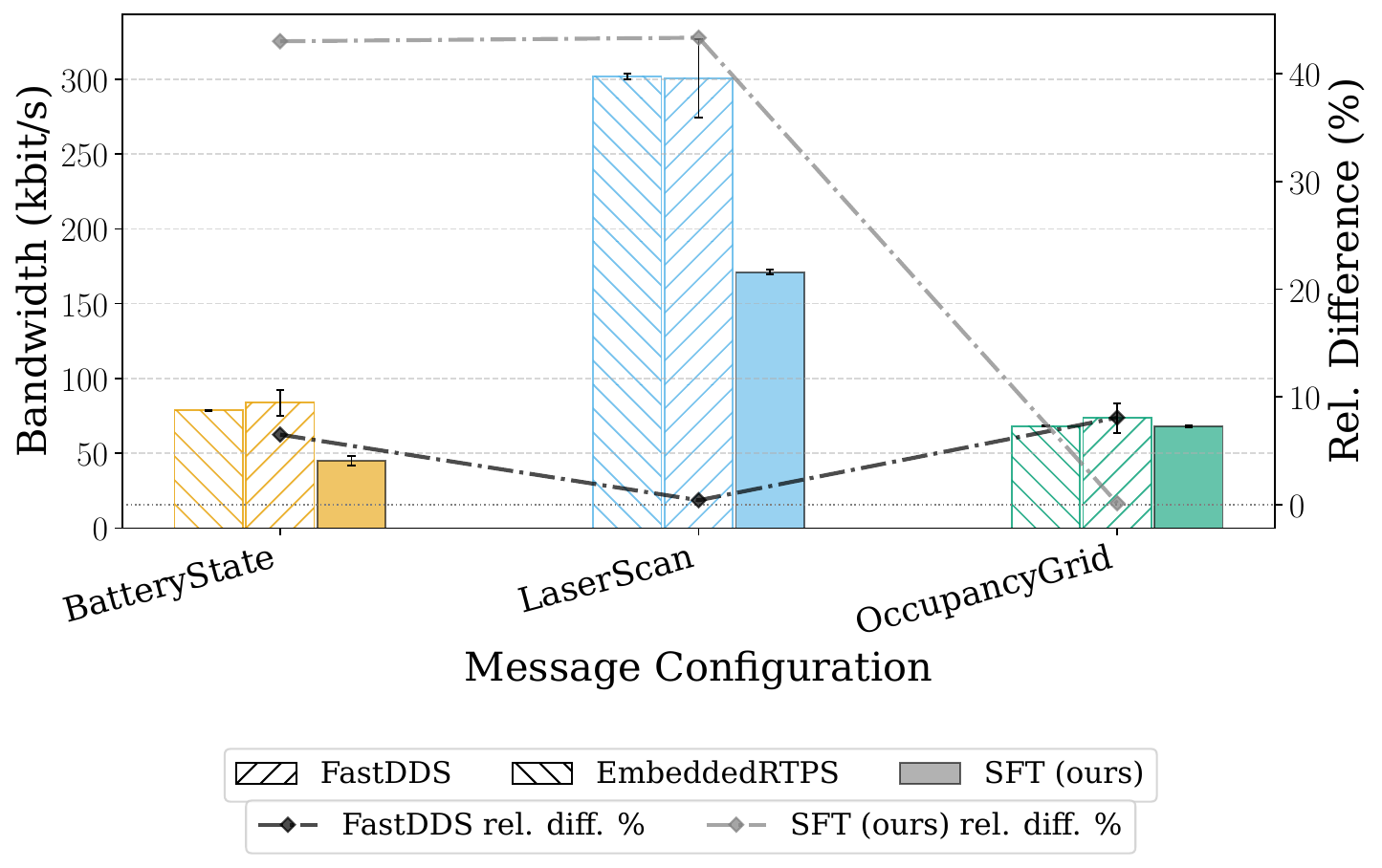}
    \caption{Overview of aggregate network bandwidth usage for our \ac{Nav2} case studies for 1-1 topology. Shown are the three case studies described in Section \ref{sec:eval:nav2} with decimal \fieldmask\ representations. Bars show means, error bars indicate one standard deviation.}
    \label{fig:results:nav2}
\end{figure}

The \ac{Nav2} case studies confirm the results of the parametric study for real world workloads. 
Whenever messages contain unused fields, bandwidth usage can be reduced and overheads remain minimal and inline with previous results. 
Fig. \ref{fig:results:nav2} shows bandwidth measurements for these experiments. 
For the \texttt{LaserScan} and \texttt{BatteryState} cases, omitting unused fields reduced network traffic by \SI{43.1}{\percent} and \SI{46.5}{\percent} respectively. 
In the case of the \texttt{LaserScan}, the omission of the large intensities array explains the sizeable bandwidth reduction, while the smaller absolute reduction in case of the \texttt{BatteryState} message is due to smaller fields and baseline protocol overhead.
In scenarios in which all fields were required, such as the \texttt{OccupancyGrid} case, \ac{SFT} introduced negligible overhead.
Latency results confirmed the parametric study findings: \ac{SFT} introduced no measurable latency overhead over EmbeddedRTPS across all three case studies.
Overall, \ac{SFT}'s selective serialization achieves bandwidth reductions with no measurable latency overhead and minimal resource overhead, providing an effective optimization without manual adaptation of interfaces.

\subsection{PointCloud2 Case Study}
\label{sec:eval:pc2}
To evaluate \ac{SFT} with heterogeneous field sizes and distinct \fieldRequirements\ on a single topic, we consider the system from Section~\ref{sec:concept:example} with three receivers and \fieldRequirements\ following the example.

\subsubsection{Results}
\begin{figure}[bp]
    \centering
    Pointcloud2 Case Studies Bandwidth Measurement\par\medskip
    \includegraphics[width=0.47\textwidth]{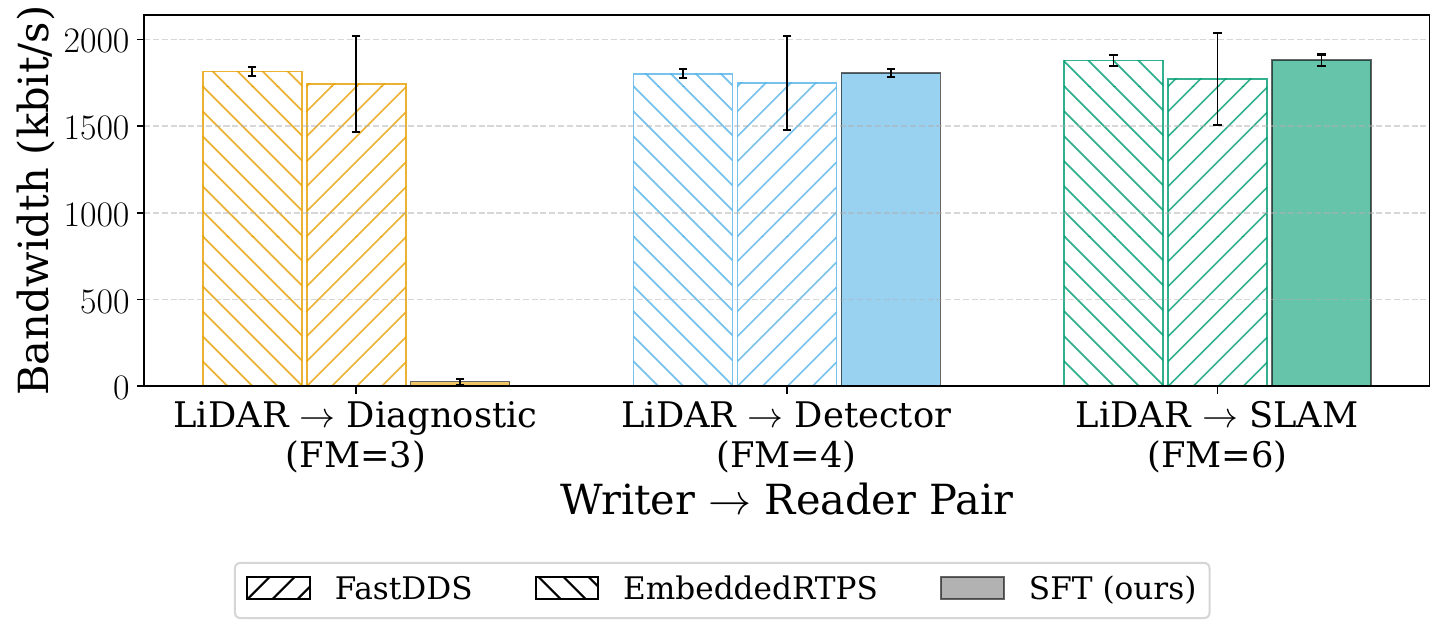}
    \caption{Overview of per-sender-receiver network bandwidth usage for our PointCloud 2 case studies following the example outlined in Section \ref{sec:concept:example}. Bars show means, error bars indicate one standard deviation.}
    \label{fig:results:pc2}
\end{figure}

Fig. \ref{fig:results:pc2} shows per-sender-receiver pair bandwidth measurements.
\ac{SFT} reduced aggregate bandwidth by \SI{32.9}{\percent} compared to EmbeddedRTPS.
This reduction was achieved because the diagnostic node, which requires only two scalar fields, was able to omit the \SI{16}{\kilo\byte} \texttt{data} field.
The object detector and SLAM node achieved insignificant reductions by omitting single scalar fields, respectively.
This case study demonstrates that \ac{SFT} serves receivers with widely different requirements from a single topic, with savings determined by the size of omitted fields.

\section{Conclusion}
\label{sec:conclusion}
This work introduces \emph{Selective Field Transmission} (SFT), a middleware-level mechanism that decouples transmission behavior from static \messageTypes\ in \ac{DDS} systems.
By transmitting only the fields actually required by each receiver, \ac{SFT} preserves interface compatibility while reducing bandwidth usage without measurable per-message latency overhead.
In our parametric study, \ac{SFT} achieved mean bandwidth reductions of \SI{19.7}{\percent}, with mean reductions of \SI{49.1}{\percent} at $FRP = \SI{50}{\percent}$.
The \ac{Nav2} and PointCloud2 case studies confirmed these results for real message structures with heterogeneous field sizes, with per-receiver reductions of up to \SI{46.5}{\percent}, while full-mask scenarios introduced negligible overhead.
Together, these results show that \ac{SFT} dissolves the trade-off between standard and project-specific \messageTypes\ in practice, allowing developers to keep reusable \messageTypes\ without paying the bandwidth cost of unused fields.
Current limitations include top-level field granularity and potentially incomplete \fieldRequirements\ under automatic requirement recording.
In future work, we plan to extend ARR with a retransmission mechanism and develop the idea of dynamic interfaces further.

\printbibliography

\end{document}